\documentclass[11pt]{revtex4}   
\usepackage{amssymb,epsf}
\usepackage{latexsym}

\begin{document}

\title{Interacting agegraphic tachyon model of dark energy}

\author{Ahmad Sheykhi \footnote{
sheykhi@mail.uk.ac.ir}}
\address{Department of Physics, Shahid Bahonar University, P.O. Box 76175, Kerman, Iran\\
         Research Institute for Astronomy and Astrophysics of Maragha (RIAAM), Maragha,
         Iran}
\begin{abstract}
Scalar-field dark energy models like tachyon are often regarded as
an effective description of an underlying theory of dark energy.
In this Letter, we implement the interacting agegraphic dark
energy models with tachyon field. We demonstrate that the
interacting agegraphic evolution of the universe can be described
completely by a single tachyon scalar field. We thus reconstruct
the potential as well as the dynamics of the tachyon field
according to the evolutionary behavior of interacting agegraphic
dark energy.
\end{abstract}
\maketitle

\section{Introduction\label{Int}}
A great variety  of cosmological observations, direct and
indirect, reveal that our universe is currently experiencing a
phase of accelerated expansion \cite{Rie}. A component which
causes cosmic acceleration is usually dubbed dark energy which
constitute a major puzzle of modern cosmology. The most obvious
theoretical candidate of dark energy is the cosmological constant.
Though, it suffers the so-called \textit{fine-tuning} and
\textit{cosmic-coincidence} problems. Among different candidates
for probing the nature of dark energy, the holographic dark energy
model arose a lot of enthusiasm recently
\cite{Coh,Li,Huang,Hsu,HDE,Setare1}. This model is motivated from
the holographic hypothesis \cite{Suss1} and has been tested and
constrained by various astronomical observations \cite{Xin,Feng}.
However, there are some difficulties in holographic dark energy
model. Choosing the event horizon of the universe as the length
scale, the holographic dark energy gives the observation value of
dark energy in the universe and can drive the universe to an
accelerated expansion phase. But an obvious drawback concerning
causality appears in this proposal. Event horizon is a global
concept of spacetime; existence of event horizon of the universe
depends on future evolution of the universe; and event horizon
exists only for universe with forever accelerated expansion. In
addition, more recently, it has been argued that this proposal
might be in contradiction to the age of some old high redshift
objects, unless a lower Hubble parameter is considered
\cite{Wei0}.

An interesting proposal to explore the nature of dark energy
within the framework of quantum gravity is a so-called agegraphic
dark energy (ADE). This model takes into account the Heisenberg
uncertainty relation of quantum mechanics together with the
gravitational effect in general relativity. The ADE model assumes
that the observed dark energy comes from the spacetime and matter
field fluctuations in the universe \cite{Cai1,Wei2,Wei1}. Since in
ADE model the age of the universe is chosen as the length measure,
instead of the horizon distance, the causality problem in the
holographic dark energy is avoided. The agegraphic models of dark
energy  have been examined and constrained by various astronomical
observations \cite{age,shey1,Setare2}. Although going along a
fundamental theory such as quantum gravity may provide a hopeful
way towards understanding the nature of dark energy, it is hard to
believe that the physical foundation of ADE is convincing enough.
Indeed, it is fair to say that almost all dynamical dark energy
models are settled at the phenomenological level, neither
holographic dark energy model nor ADE model is exception. Though,
under such circumstances, the models of holographic and ADE, to
some extent, still have some advantage comparing to other
dynamical dark energy models because at least they originate from
some fundamental principles in quantum gravity.

On the other hand, among the various candidates to explain the
accelerated expansion, the rolling tachyon condensates in a class
of string theories may have interesting cosmological consequences.
The tachyon is an unstable field which has became important in
string theory through its role in the Dirac-Born-Infeld action
which is used to describe the D-brane action \cite{Sen1,Sen2}. It
has been shown \cite{Sen3} that the decay of D-branes produces a
pressureless gas with finite energy density that resembles
classical dust. The effective Lagrangian for the tachyon field is
described by
\begin{eqnarray}
 L=-V(\phi)\sqrt{1-g^{\mu\nu}\partial_\mu \phi \partial_\nu \phi},
 \end{eqnarray}
where $V(\phi)$ is the tachyon potential. The corresponding energy
momentum tensor for the tachyon field can be written in a perfect
fluid form
\begin{eqnarray}
 T_{\mu\nu}=(\rho_\phi+p_\phi)u_{\mu} u_\nu-p_\phi g_{\mu\nu},
 \end{eqnarray}
where $\rho_\phi$ and $p_\phi$ are, respectively, the energy
density and pressure of the tachyon and the velocity $u_\mu$ is
\begin{eqnarray}
 u_\mu=\frac{\partial_\mu \phi}{\sqrt{\partial_\nu \phi \partial^\nu
 \phi}}.
 \end{eqnarray}
A rolling tachyon has an interesting equation of state whose
parameter smoothly interpolates between $-1$ and $0$ \cite{Gib1}.
Thus, tachyon can be realized as a suitable candidate for the
inflation at high energy \cite{Maz1} as well as a source of dark
energy depending on the form of the tachyon potential \cite{Padm}.
Therefore it becomes meaningful to reconstruct tachyon potential
$V(\phi)$ from some dark energy models possessing some significant
features of the quantum gravity theory, such as holographic and
ADE models. It was demonstrated that dark energy driven by
tachyon, decays to cold dark matter in the late accelerated
universe and this phenomenon yields a solution to cosmic
coincidence problem \cite{Sri}. The investigations on the
reconstruction of the tachyon potential $V(\phi)$ in the framework
of holographic dark energy have been carried out in
\cite{Setare4}. In the absence of the interaction between ADE and
dark matter, the connection between tachyon field and the new ADE
model has also been established in \cite{agetach}.

In the present Letter, we would like to extend the study to the
case where both components- the pressureless dark matter and the
ADE- do not conserve separately but interact with each other.
Given the unknown nature of both dark matter and dark energy there
is nothing in principle against their mutual interaction and it
seems very special that these two major components in the universe
are entirely independent \cite{Setare3,wang1,shey2}. We shall
establish a correspondence between the interacting ADE scenarios
and the tachyon scalar field in a non-flat universe. Although it
is believed that our universe is flat, a contribution to the
Friedmann equation from spatial curvature is still possible if the
number of e-foldings is not very large \cite{Huang}. Besides, some
experimental data has implied that our universe is not a perfectly
flat universe and recent papers have favored the universe with
spatial curvature \cite{spe}. We suggest the agegraphic
description of the tachyon dark energy in a universe with spacial
curvature and reconstruct the potential and the dynamics of the
tachyon scalar field which describe the tachyon cosmology. The
plan of the work is as follows. In the next section we associate
the original ADE with the tachyon field. In section \ref{NEW}, we
establish the correspondence between the new model of interacting
ADE and the tachyon dark energy. The last section is devoted to
conclusions.

\section{Tachyon reconstruction of the ORIGINAL ADE  \label{ORI}}
We consider the Friedmann-Robertson-Walker (FRW) universe which is
described by the line element
\begin{eqnarray}
 ds^2=dt^2-a^2(t)\left(\frac{dr^2}{1-kr^2}+r^2d\Omega^2\right),\label{metric}
 \end{eqnarray}
where $a(t)$ is the scale factor, and $k$ is the curvature
parameter with $k = -1, 0, 1$ corresponding to open, flat, and
closed universes, respectively. A closed universe with a small
positive curvature ($\Omega_k\simeq0.01$) is compatible with
observations \cite{spe}. The first Friedmann equation takes the
form
\begin{eqnarray}\label{Fried}
H^2+\frac{k}{a^2}=\frac{1}{3m_p^2} \left( \rho_m+\rho_D \right).
\end{eqnarray}
We define, as usual, the fractional energy densities such as
\begin{eqnarray}\label{Omega}
\Omega_m=\frac{\rho_m}{3m_p^2H^2}, \hspace{0.5cm}
\Omega_D=\frac{\rho_D}{3m_p^2H^2},\hspace{0.5cm}
\Omega_k=\frac{k}{H^2 a^2},
\end{eqnarray}
thus, the Friedmann equation can be written
\begin{eqnarray}\label{Fried2}
\Omega_m+\Omega_D=1+\Omega_k.
\end{eqnarray}
We adopt the viewpoint that the scalar field models of dark energy
are effective theories of an underlying theory of dark energy. The
energy density and pressure for the tachyon scalar field can be
written as
\begin{eqnarray}\label{rhophi}
\rho_\phi&=&-T^0 _0=\frac{V(\phi)}{\sqrt{1-\dot{\phi}^2}},\\
p_\phi&=&T^i _i=-V(\phi)\sqrt{1-\dot{\phi}^2}. \label{pphi}
\end{eqnarray}
Consequently the equation of state of the tachyon is given by
\begin{eqnarray}\label{wphi}
w_\phi=\frac{p_\phi}{\rho_\phi}=\dot{\phi}^2-1.
\end{eqnarray}
From Eq. (\ref{wphi}) we see that irrespective of the steepness of
the tachyon potential, we have always $-1<w_\phi<0$. This implies
that the tachyon field cannot realize the equation of state
crossing $-1$.  Next we intend to implement the interacting
original ADE models with tachyon scalar field. Let us first review
the origin of the ADE model. Following the line of quantum
fluctuations of spacetime, Karolyhazy et al. \cite{Kar1} argued
that the distance $t$ in Minkowski spacetime cannot be known to a
better accuracy than $\delta{t}=\beta t_{p}^{2/3}t^{1/3}$ where
$\beta$ is a dimensionless constant of order unity. Based on
Karolyhazy relation, Maziashvili discussed that the energy density
of metric fluctuations of the Minkowski spacetime is given by
\cite{Maz}
\begin{equation}\label{rho0}
\rho_{D} \sim \frac{1}{t_{p}^2 t^2} \sim \frac{m^2_p}{t^2},
\end{equation}
where $t_{p}$ is the reduced Planck time and $t$ is a proper time
scale. In the original ADE model Cai \cite{Cai1} proposed the dark
energy density of the form (\ref{rho0}) where $t$ is chosen to be
the age of the universe
\begin{equation}
T=\int_0^a{\frac{da}{Ha}},
\end{equation}
Thus, he wrote down the energy density of the original ADE as
\cite{Cai1}
\begin{equation}\label{rho1}
\rho_{D}= \frac{3n^2 m_{p}^2}{T^2},
\end{equation}
where the numerical factor $3n^2$ is introduced to parameterize
some uncertainties, such as the species of quantum fields in the
universe, the effect of curved space-time, and so on. The dark
energy density (\ref{rho1}) has the same form as the holographic
dark energy, but  the length measure is chosen to be the age of
the universe instead of the horizon radius of the universe. Thus
the causality problem in the holographic dark energy is avoided.
Combining Eqs. (\ref{Omega}) and (\ref{rho1}), we get
\begin{eqnarray}\label{Omegaq}
\Omega_D=\frac{n^2}{H^2T^2}.
\end{eqnarray}
The total energy density is $\rho=\rho_{m}+\rho_{D}$, where
$\rho_{m}$ and $\rho_{D}$ are the energy density of dark matter
and dark energy, respectively. The total energy density satisfies
a conservation law
\begin{equation}\label{cons}
\dot{\rho}+3H(\rho+p)=0.
\end{equation}
However, since we consider the interaction between dark matter and
dark energy, $\rho_{m}$ and $\rho_{D}$ do not conserve separately;
they must rather enter the energy balances
\begin{eqnarray}
&&\dot{\rho}_m+3H\rho_m=Q, \label{consm}
\\&& \dot{\rho}_D+3H\rho_D(1+w_D)=-Q.\label{consq}
\end{eqnarray}
Here $w_D$ is the equation of state parameter of ADE and $Q$
denotes the interaction term and can be taken as $Q =3b^2 H\rho$
with $b^2$  being a coupling constant \cite{Pav1}. Taking the
derivative with respect to the cosmic time of Eq. (\ref{rho1}) and
using Eq. (\ref{Omegaq}) we get
\begin{eqnarray}\label{rhodot}
\dot{\rho}_D=-2H\frac{\sqrt{\Omega_D}}{n}\rho_D.
\end{eqnarray}
Inserting this relation into Eq. (\ref{consq}), we obtain the
equation of state parameter of the original ADE in non-flat
universe
\begin{eqnarray}\label{wq}
w_D=-1+\frac{2}{3n}\sqrt{\Omega_D}-\frac{b^2}{\Omega_D}
(1+\Omega_k).
\end{eqnarray}
Differentiating Eq. (\ref{Omegaq}) and using relation
${\dot{\Omega}_D}={\Omega'_D}H$, we reach
\begin{eqnarray}\label{Omegaq2}
{\Omega'_D}=\Omega_D\left(-2\frac{\dot{H}}{H^2}-\frac{2}{n
}\sqrt{\Omega_D}\right),
\end{eqnarray}
where the dot and the prime stand for the derivative with respect
to the cosmic time and the derivative with respect to $x=\ln{a}$,
respectively. Taking the derivative of both side of the Friedman
equation (\ref{Fried}) with respect to the cosmic time, and using
Eqs. (\ref{Fried2}), (\ref{rho1}), (\ref{Omegaq}) and
(\ref{consm}), it is easy to show that
\begin{eqnarray}\label{Hdot}
\frac{\dot{H}}{H^2}=-\frac{3}{2}(1-\Omega_D)-\frac{\Omega^{3/2}_D}{n}-\frac{\Omega_k}{2}
+\frac{3}{2}b^2(1+\Omega_k).
\end{eqnarray}
Substituting this relation into Eq. (\ref{Omegaq2}), we obtain the
equation of motion for the original ADE
\begin{eqnarray}\label{Omegaq3}
{\Omega'_D}&=&\Omega_D\left[(1-\Omega_D)\left(3-\frac{2}{n}\sqrt{\Omega_D}\right)
-3b^2(1+\Omega_k)+\Omega_k\right].
\end{eqnarray}
From the first Friedmann equation (\ref{Fried}) sa well as Eqs.
(\ref{Fried2}), (\ref{consm}) and  (\ref{consq}), we obtain
\begin{eqnarray}\label{HH0}
H=H_0\sqrt{\frac{1+\Omega_{k_0}}{1+\Omega_k}}\
\exp\left[-\frac{3}{2}\int_{a_0}^{a}{(1+w_D)\frac{da}{a}}\right].
\end{eqnarray}
Now we suggest a correspondence between the original ADE and
tachyon scalar field namely, we identify $\rho_\phi$ with
$\rho_D$. Using relation $\rho_\phi=\rho_D={3m_p^2H^2}\Omega_D$
and Eqs. (\ref{rhophi}), (\ref{wphi}) and (\ref{wq}), we can find
\begin{eqnarray}\label{vphi2}
V(\phi)&=&\rho_\phi\sqrt{1-\dot{\phi}^2}=3m^2_pH^2 \Omega_D\left(1-\frac{2}{3n}\sqrt{\Omega_D}+\frac{b^2}{\Omega_D}(1+\Omega_k)\right)^{1/2},\\
\dot{\phi}&=&=\sqrt{1+w_D}=\left(\frac{2}{3n}\sqrt{\Omega_D}-\frac{b^2}{\Omega_D}
(1+\Omega_k)\right)^{1/2}.\label{dotphi2}
\end{eqnarray}
Using relation $\dot{\phi}=H{\phi'}$, we get
\begin{eqnarray}\label{primephi}
{\phi'}&=&H^{-1}\left(\frac{2}{3n}\sqrt{\Omega_D}-\frac{b^2}{\Omega_D}
(1+\Omega_k)\right)^{1/2}.
\end{eqnarray}
Consequently, we can easily obtain the evolutionary form of the
 tachyon field by integrating the above equation
\begin{eqnarray}\label{phi}
\phi(a)-\phi(a_0)=\int_{a_0}^{a}{\frac {1}{H
a}\sqrt{\frac{2}{3n}\sqrt{\Omega_D}-\frac{b^2}{\Omega_D}
(1+\Omega_k)}\ da},
\end{eqnarray}
where $a_0$  is the  value of the scale factor at the present time
$t_0$, $H$ is given by Eq. (\ref{HH0}) and $\Omega_D$ can be
extracted from Eq. (\ref{Omegaq3}). The above equation can also be
written in the following form
\begin{eqnarray}\label{phit}
\phi(t)-\phi(t_0)=\int_{t_0}^{t}{\sqrt{\frac{2}{3n}\sqrt{\Omega_D}-\frac{b^2}{\Omega_D}
(1+\Omega_k)}\ dt'}.
\end{eqnarray}
Therefore, we have established an interacting agegraphic tachyon
dark energy model and reconstructed the potential and the dynamics
of the tachyon field. It is worth noting that if one omits
$\Omega_D$ between Eqs. (\ref{vphi2}) and  (\ref{phi}), one can
obtain $V=V(\phi)$. Unfortunately, this cannot be done
analytically for the above general solutions. Let us consider, as
an example, the matter-dominated epoch where $a\ll1$ and $\Omega_D
\ll1$. In this case Eq. (\ref{Omegaq3}) with $\Omega_k \ll 1$
approximately becomes
\begin{eqnarray}\label{Omegaq32}
\frac{d\Omega_D}{da}\simeq \frac{\Omega_D}{a}
\left(3-\frac{2}{n}\sqrt{\Omega_D}-3b^2\right),
\end{eqnarray}
Solving the above equation we find
\begin{eqnarray}\label{Omegaq33}
 \Omega_D =\frac{9n^2}{4} (1-b^2)^2.
\end{eqnarray}
Substituting this relation into Eq. (\ref{wq}), we obtain
\begin{eqnarray}\label{wqq2}
w_D=-b^2 \left(1+\frac{4}{9n^2(1-b^2)^2}\right).
\end{eqnarray}
In this case for $\Omega_k \ll 1$, Eq. (\ref{HH0}) can be
integrated. The result is
\begin{eqnarray}\label{H3}
H=H_0 a^{-3(1+w_D)/2}.
\end{eqnarray}
Combining Eqs. (\ref{Omegaq33}) and (\ref{H3}) with (\ref{phi}) we
find
\begin{eqnarray}\label{phi4}
\phi=\frac{2a^{3(1+w_D)/2}}{3H_0 \sqrt{1+w_D}},
\end{eqnarray}
up to a constant of integration.  From this equation we get
\begin{eqnarray}\label{a1}
a= \left(\frac{9(1+w_D)\phi^2
H_0^2}{4}\right)^{\frac{1}{3(1+w_D)}}.
\end{eqnarray}
Finally, combining Eqs. (\ref{Omegaq33}), (\ref{wqq2}), (\ref{a1})
with Eq. (\ref{vphi2}) we reach
\begin{eqnarray}\label{vv}
V(\phi)= \,{\frac {{{\it -9 m_p}}^{2} \left( -1+{b}^{2} \right)
^{3}{n}^{3}b \sqrt
{-9\,{n}^{2}+18\,{n}^{2}{b}^{2}-9\,{n}^{2}{b}^{4}-4}}{ \left( -9
\,{n}^{2}+27\,{n}^{2}{b}^{2}-27\,{n}^{2}{b}^{4}+9\,{n}^{2}{b}^{6}+4\,{
b}^{2} \right) {\phi}^{2}}}.
\end{eqnarray}
\section{Tachyon reconstruction of the NEW ADE  \label{NEW}}
To avoid some internal inconsistencies in the original ADE model,
the so-called ``new agegraphic dark energy" was proposed, where
the time scale is chosen to be the conformal time $\eta$ instead
of the age of the universe \cite{Wei2}. The new ADE contains some
new features different from the original ADE and overcome some
unsatisfactory points. For instance, the original ADE suffers from
the difficulty to describe the matter-dominated epoch while the
new ADE resolved this issue \cite{Wei2}. The energy density of the
new ADE can be written
\begin{equation}\label{rho1new}
\rho_{D}= \frac{3n^2 m_{p}^2}{\eta^2},
\end{equation}
where the conformal time $\eta$ is given by
\begin{equation}
\eta=\int{\frac{dt}{a}}=\int_0^a{\frac{da}{Ha^2}}.
\end{equation}
The fractional energy density of the new ADE is now expressed as
\begin{eqnarray}\label{Omegaqnew}
\Omega_D=\frac{n^2}{H^2\eta^2}.
\end{eqnarray}
Taking the  derivative with respect to the cosmic time of Eq.
(\ref{rho1new}) and using Eq. (\ref{Omegaqnew}) we get
\begin{eqnarray}\label{rhodotnew}
\dot{\rho}_D=-2H\frac{\sqrt{\Omega_D}}{na}\rho_D.
\end{eqnarray}
Inserting this relation into Eq. (\ref{consq}) we obtain the
equation of state parameter of the new ADE
\begin{eqnarray}\label{wqnew}
w_D=-1+\frac{2}{3na}\sqrt{\Omega_D}-\frac{b^2}{\Omega_D}
(1+\Omega_k).
\end{eqnarray}
The evolution behavior of the new ADE is now given by
\begin{eqnarray}\label{Omegaq3new}
{\Omega'_D}&=&\Omega_D\left[(1-\Omega_D)\left(3-\frac{2}{na}\sqrt{\Omega_D}\right)
-3b^2(1+\Omega_k)+\Omega_k\right].
\end{eqnarray}
Next, we reconstruct the new agegraphic tachyon dark energy model,
connecting the tachyon scalar field with the new ADE. Using Eqs.
(\ref{Omegaqnew}) and (\ref{wqnew}) one can easily show that the
tachyon potential and kinetic energy term take the following form
\begin{eqnarray}\label{vphi2new}
V(\phi)&=&3m^2_pH^2 \Omega_D\left(1-\frac{2}{3na}\sqrt{\Omega_D}+\frac{b^2}{\Omega_D}(1+\Omega_k)\right)^{1/2},\\
\dot{\phi}&=&\left(\frac{2}{3na}\sqrt{\Omega_D}-\frac{b^2}{\Omega_D}
(1+\Omega_k)\right)^{1/2}.\label{dotphi2new}
\end{eqnarray}
We can also rewrite Eq. (\ref{dotphi2new}) as
\begin{eqnarray}\label{primephinew}
{\phi'}&=&H^{-1}\left(\frac{2}{3na}\sqrt{\Omega_D}-\frac{b^2}{\Omega_D}
(1+\Omega_k)\right)^{1/2}.
\end{eqnarray}
Therefore the evolution behavior of the tachyon field can be
obtained by integrating the above equation
\begin{eqnarray}\label{phinew}
\phi(a)-\phi(a_0)=\int_{a_0}^{a}{\frac {1}{H
a}\sqrt{\frac{2}{3na}\sqrt{\Omega_D}-\frac{b^2}{\Omega_D}
(1+\Omega_k)}\ da},
\end{eqnarray}
or in another way
\begin{eqnarray}\label{phitnew}
\phi(t)-\phi(t_0)=\int_{t_0}^{t}{\sqrt{\frac{2}{3na}\sqrt{\Omega_D}-\frac{b^2}{\Omega_D}
(1+\Omega_k)}\ dt'}.
\end{eqnarray}
where $\Omega_D$ is now given by Eq. (\ref{Omegaq3new}). In this
way we connect the interacting new ADE with a tachyon field and
reconstruct the potential and the dynamics of the tachyon field
which describe tachyon cosmology.
\section{Conclusions\label{CONC}}
Among the various candidates to play the role of the dark energy,
tachyon has emerged as a possible source of dark energy for a
particular class of potentials \cite{Padm}. In this Letter, we
have associated the interacting ADE models with a tachyon field
which describe the tachyon cosmology in a non-flat universe. The
ADE models take into account the Heisenberg uncertainty relation
of quantum mechanics together with the gravitational effect in
general relativity. These models assume that the observed dark
energy comes from the spacetime and matter field fluctuations in
the universe. Therefore, agegraphic scenarios may possess some
significant features of an underlying theory of dark energy. We
have demonstrated that the agegraphic evolution of the universe
can be described completely by a tachyon scalar field in a certain
way. We have adopted the viewpoint that the scalar field models of
dark energy are effective theories of an underlying theory of dark
energy. Thus, we should be capable of using the scalar field model
to mimic the evolving behavior of the interacting ADE and
reconstructing this scalar field model. We have reconstructed the
potential and the dynamics of the tachyon scalar field according
to the evolutionary behavior of the interacting agegraphic dark
energy models.

\acknowledgments{I thank the anonymous referee for constructive
comments. This work has been supported by Research Institute for
Astronomy and Astrophysics of Maragha.}

\end{document}